\pdfoutput=1

\documentclass[twocolumn,english]{revtex4}
\usepackage{times,amssymb,amsmath,graphicx}
\usepackage[bookmarks=false,pdffitwindow=false,pdfstartview={FitH}]{hyperref}
\usepackage[T1]{fontenc}
\usepackage{babel}
\usepackage{color}

\begin{document}


\title{
  Sub-Sharvin conductance and enhanced shot noise in doped graphene 
}

\author{Adam Rycerz}
\affiliation{Institute for Theoretical Physics,
  Jagiellonian University, \L{}ojasiewicza 11, PL--30348 Krak\'{o}w, Poland}

\author{Piotr Witkowski}
\affiliation{Institute for Theoretical Physics,
  Jagiellonian University, \L{}ojasiewicza 11, PL--30348 Krak\'{o}w, Poland}

\date{September 24, 2021}

\begin{abstract}
  Ideal Sharvin contact in a~multimode regime shows the conductance 
  $G\approx{}G_{\rm Sharvin}=g_0k_F{}W/\pi$ (with $g_0$ the conductance quantum,
  $k_F$ the Fermi momentum, and $W$ the contact width) accompanied by 
  strongly suppressed shot-noise quantified by small Fano factor
  $F\approx{}0$. 
  For ballistic graphene away from the charge-neutrality point the sub-Sharvin
  transport occurs, characterised by suppressed conductance
  $G\approx{}(\pi/4)\,G_{\rm Sharvin}$ and enhanced shot noise $F\approx{}1/8$.
  All these results can be derived from a~basic model of quantum scattering, 
  involving assumptions of infinite height and perfectly rectangular shape
  of the potential barrier in the sample. 
  Here we have carried out the numerical analysis of the scattering on
  a family of smooth barriers of finite height interpolating between
  parabollic and rectangular shapes.
  We find that tuning the barrier shape one can modify the asymmetry
  between electron- and hole-doped systems.
  For electronic dopings, the system crosses from Sharvin to sub-Sharvin
  transport regime (indicated by both the conductance and the Fano factor)
  as the potential becomes closer to the rectangular shape.
  In contrast, for hole dopings, the conductivity is strongly suppressed
  when the barrier is parabolic and slowly converges to
  $G\approx{}(\pi/4)\,G_{\rm Sharvin}$ 
  as the potential evolves towards rectangular shape. 
  In such a~case the Sharvin transport regime is inaccessible,
  shot noise is generically enhanced (with much slower convergence
  to $F\approx{}1/8$) comparing to the electron-doped case, 
  and aperiodic oscillations of both $G$ and $F$  are prominent 
  due to the formation of quasibound states. 
\end{abstract}

\maketitle

\section{Introduction}
Soon after the isolation of monolayer graphene \cite{Nov04} experimental and
theoretical physicists have reexamined classical effects from
mesoscopic phycics \cite{Nov05,Zha05,Rus10,Lin08,Tom11,Pal12,Ryc13,Hua18,Zen19}.
In ballistic graphene ribbons \cite{Lin08} or constrictions \cite{Tom11},
showing conductance quantization, electrical conductance approaches
the fundamental upper bound given by the Sharvin formula \cite{Sha65,Bee91}.
A~few years ago, ultraclean graphene samples exhibiting a~viscous
charge flow due to electron-electron interactions \cite{Luc18} allowed
to detect the conductance exceeding the Sharvin bound \cite{Guo17,Kum17}. 

Since the spectrum of excitations in graphene consists of two conical bands
and is described by a~two-dimensional analog of the relativistic Dirac
equation \cite{Mcc56,Sem84,Div84}, several novel effects can be identified
even at low temperatures, where interactions become negligible and
ballistic (or Landauer-B\"{u}ttiker) transport regime
is restored \cite{Kat06,Two06,Mia07,Dan08,Ryc10,Ric15,Kum18}.
For instance, the phenomenon of Klein tunneling \cite{Kat12} manifests
itself via the universal conductivity ($\sigma_0=4e^2/\pi{}h$, with
the electron charge $-e$ and the Planck constant $h$) and the so-called
{\em pseudodiffusive} shot noise power (quantified by the Fano factor $F=1/3$)
\cite{Two06,Mia07,Dan08} provided that carrier concentration is close to the
charge-neutrality point.

Although several features of ballistic graphene may also be observed in
other two-dimensional systems \cite{Eza13,Eza15,Rze20} universal conductivity
seems to be the unique feature of graphene, having no direct analog even in
bilayer graphene \cite{Sus20}. 

Remarkably, in the universal-conductivity range (for a~rectangular sample,
it is further required that the width $W\gg{}L$ with $L$ being the length,
see Ref.\ \cite{Ryc09}) 
the conductance is enhanced due to the transport via evanescent waves,
while the shot noise is suppressed. 
However, tuning the carrier concentration away from the charge-neutrality
point, results in the Fano factor approaching values in a~range of
$F\approx{}0.10\div{}0.15$ \cite{Two06,Dan08}, being significantly greater
then $F\approx{}0$ expected for a~ballistic system.
The conductance is difficult to be determined experimentally due to
resistances of contacts, but the simple analytical discussion leads to
the value reduced by a~factor of $\pi/4$ compared to the Sharvin formula
in the high-doping limit \cite{Par21,Ryc21}. (Same analysis leads to the
Fano factor converging to $F\rightarrow{}1/8$.) 

\begin{figure}[!t]
  \includegraphics[width=0.9\linewidth]{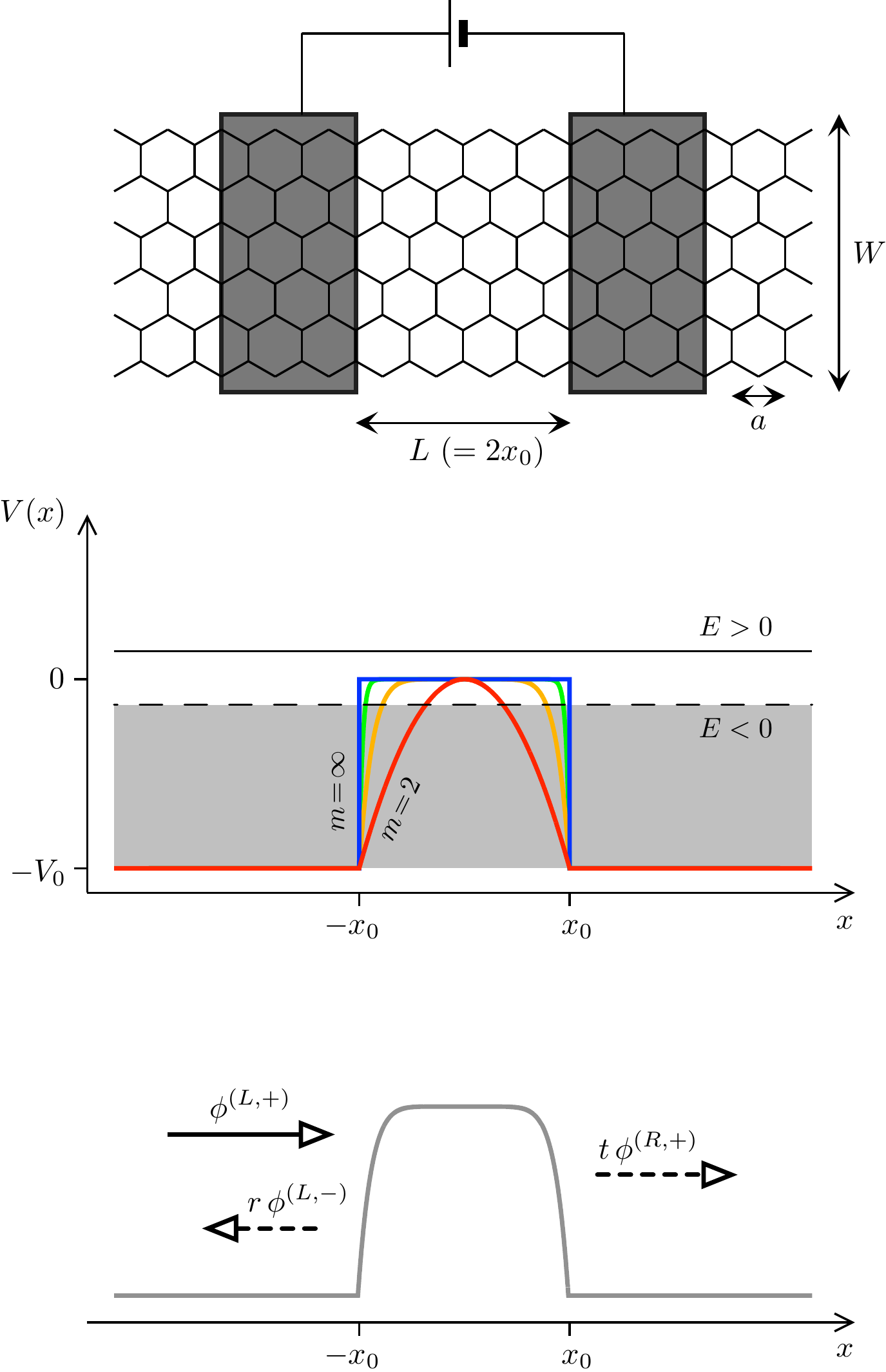}
  \caption{ \label{setup:fig}
    Top: Schematic of a~graphene strip of width $W$, contacted by
    two electrodes (dark areas) at a~distance $L$. 
    A voltage source drives a current through the strip.
    A separate gate electrode (not shown) allows to tune the carrier
    concentration around the neutrality point. The lattice parameter
    $a=0.246\,$nm is also shown.
    Middle: Electrostatic potential profiles given by Eq.\ (\ref{v0mpot})
    with $m=2,8,32,$ and $m=\infty$ (i.e., the rectangular barrier).
    The Fermi energy $E$ is defined with respect to the top of a~barrier.
    $E>0$ corresponds to unipolar n-n-n doping in the device;
    for $E<0$, a~n-p-n structure is formed.
    Bottom: A~symbolic representation of the incident and reflected
    waves in left electrode ($x<-x_0$) and the transmitted wave in right
    electrode ($x>x_0$) with the amplitudes $r$ and $t$. 
  }
\end{figure}

The purpose of this work is to investigate numerically, how the conductance
and Fano factor for a~ballistic graphene sample behave as functions of
doping supposing that the electrostatic potential barrier is smooth
(i.e., potential varies slowly on the scale of atomic separation).
Similar problems were addressed previously \cite{Par21,Sil07,Che06,Cay09},
but here we focus on the effects of the potential profile, which is 
gradually tuned from a~parabolic to a~rectangular shape
(see Fig.\ \ref{setup:fig}), on the selected transport properties. 

The paper is organized as follows. In Sec.\ \ref{modmets}, we present
the details of our numerical approach. The key results for a~rectangular
barrier are summarized in Sec.\ \ref{rectbar}. 
Our main results, concerning the conductance and Fano factor for smooth
potentials are presented in Sec.\ \ref{resudis}. 
The conclusions are given in Sec.\ \ref{conclus}.

\section{Model and methods}
\label{modmets}

We start from the scattering problem for massless Dirac fermions
in graphene at the energy $E$, in case when the electrostatic potential
energy depends only on the $x$ coordinate, i.e., $V\equiv{}V(x)$.
The wave equation can be written as
\begin{equation}
  \left[
    v_F\,\mbox{\boldmath$p$}\cdot\mbox{\boldmath$\sigma$} + V(x)
  \right]
  \Psi=E\Psi, 
\end{equation}
where $v_F=\sqrt{3}\,t_0a/(2\hbar)\approx{}10^6\,$m$/$s is the
energy-independent Fermi velocity in graphene (with $t_0=2.7\,$eV
the nearest-neighbor hopping integral and $a=0.246$ the lattice parameter),
$\mbox{\boldmath$p$}=(p_x,p_y)$ is the in-plane momentum operator with
$p_j=-i\hbar{}\partial_j$, and $\mbox{\boldmath$\sigma$}=(\sigma_x,\sigma_y)$
with $\sigma_j$ being the Pauli matrices. 
Taking the wavefunction in a~form $\Psi = \phi(x)e^{ik_yy}$, with
$\phi(x) = (\phi_a,\phi_b)^T$ and $k_y$ the transverse wavenumber,
brought us to the system of ordinary differential equations for
the spinor components
\begin{align}
  \phi_a' &= k_y\phi_a+i\frac{E-V(x)}{\hbar{}v_F}\phi_b,
  \label{phapri} \\
  \phi_b' &= i\frac{E-V(x)}{\hbar{}v_F}\phi_a-k_y\phi_b.
  \label{phbpri} 
\end{align}

In a~general case of $k_y\neq{}0$, Eqs.\ (\ref{phapri},\ref{phbpri})
need to be integrated numerically \cite{Sil07}.
If one assumes the so-called infinite-mass boundary conditions \cite{Ber87}
at $y=0$ and $y=W$, in the momentum $k_y$ gets quantized
$k_y^{(n)}=\pi{}(n+\frac{1}{2})/W$, with $n=0,1,2,\dots$, \cite{Two06}.
This form is used throughout the paper. (In the limit of $W\gg{}L$,
one can also treat $k_y$ as a~as a continuous variable.)

The electrostatic potential energy is chosen as 
\begin{equation}
  \label{v0mpot}
  V(x) = -V_0\times
  \begin{cases}
    \,\left|x/x_0\right|^m  &  \text{if }\ |x| \leqslant x_0, \\
    \,1  &  \text{if }\ |x| > x_0, 
  \end{cases}
\end{equation}
such that changing the value of $m$ tunes the potential 
from parabolic shape ($m=2$) to rectangular shape ($m\rightarrow\infty$).
Above we use a~parameter $x_0=L/2$, with $L$ the sample length. 
The potential given by Eq.\ (\ref{v0mpot}) is continuous and constant in
the leads ($x<-L/2$ or $x>L/2$). 

The basis solutions in the leads, for $E>-V_0$, are
\begin{equation}
  \label{leadspm}
  \phi^{(+)}=\left(
  \begin{matrix} 1 \\ e^{i\theta} \\ \end{matrix}
  \right)
  e^{iK_x{}x},
  \ \ \ \ 
  \phi^{(-)}=\left(
  \begin{matrix} 1 \\ -e^{-i\theta} \\ \end{matrix}
  \right)
  e^{-iK_x{}x}, 
\end{equation}
where $e^{i\theta}=(K_x+ik_y)/K$, $K =(E+V_0)/\hbar{}v_F$, and
$K_x=\sqrt{K^2-k_y^2}$.
Transverse momentum is conserved in the scattering \cite{leadfoo}, 
so the value of the quantum number $k_y$ is the same for both leads
and the sample area. 
Supposing scattering from the left direction ($x=-\infty$),
the wavefunctions in the left ($L$) and right ($R$) lead can
be written as 
\begin{equation}
  \phi_{E,k_y}^{(L)}=\phi^{(+)}+r\phi^{(-)},
  \ \ \ \ \ 
  \phi_{E,k_y}^{(R)}=t\phi^{(+)},
\end{equation}
where we have defined the reflection ($r$) and transmission ($t$)
amplitudes. 

For the sample area ($-L/2<x<L/2$) the wavefunction takes a~form
\begin{equation}
  \phi_{E,k_y}^{(c)}=A\phi^{(1)}+B\phi^{(2)}, 
\end{equation}
where $\phi^{(1)}$, $\phi^{(2)}$ denote the two linearly-independent solutions
of Eqs.\ (\ref{phapri},\ref{phbpri}), which can be obtained numerically
choosing the initial conditions, say
$\left.\phi^{(1,2)}\right|_{x=-x_0}=(1,\pm{}1)^T$, and $A$, $B$
are arbitrary complex coefficients.

The matching conditions for $x=\pm{}x_0$, namely
\begin{equation}
  \phi_{E,k_y}^{(L)}=\left.\phi_{E,k_y}^{(c)}\right|_{x=-x_0}
  \ \ \ \text{and} \ \ \ 
  \phi_{E,k_y}^{(c)}=\left.\phi_{E,k_y}^{(R)}\right|_{x=x_0},
\end{equation}
immediately leads to the linear system of equations
\begin{widetext}
\begin{equation}
  \label{lsysrt}
  \left[
    \begin{matrix}
      \phi_a^{(-)}(-x_0) & -\phi_a^{(1)}(-x_0)  & -\phi_a^{(2)}(-x_0) & 0 \\
      \phi_b^{(-)}(-x_0) & -\phi_b^{(1)}(-x_0) & -\phi_b^{(2)}(-x_0) & 0 \\
      0 &  -\phi_a^{(1)}(x_0)  & -\phi_a^{(2)}(x_0)  & \phi_a^{(+)}(x_0) \\
      0 &  -\phi_b^{(1)}(x_0)  & -\phi_b^{(2)}(x_0)  & \phi_b^{(+)}(x_0) \\
    \end{matrix}
  \right]
  \left[
    \begin{matrix}
      r \\ A \\ B \\ t \\
    \end{matrix}
  \right]
  =
  \left[
    \begin{matrix}
      -\phi_a^{(+)}(-x_0) \\
      -\phi_b^{(+)}(-x_0) \\
      0 \\
      0 \\
    \end{matrix}
  \right], 
\end{equation}
\end{widetext}
where we have explicitly written the spinor components and omitted repeating
indices $(E,k_y)$ for clarity.

Solving Eq.\ (\ref{lsysrt}), one finds the transmission amplitude $t$
for a~given $E$ and $k_y$, and the corresponding transmission probability
$T_{k_y}(E)=|t|^2$. The conductance and Fano factor follow by summing
over the modes,
\begin{equation}
  \label{gfland}
  G = g_0\sum_{n=0}^{N-1}T_n, \ \ \ \ \ \ 
  F = \frac{\sum_{n=0}^{N-1}T_n(1-T_n)}{\sum_{n=0}^{N-1}T_n}, 
\end{equation}
with $g_0=4e^2/h$ (the factor $4$ accounts for spin and valley degeneracy),
$T_n=T_{k_y}(E)$ for $k_y^{(n)}=\pi{}(n+\frac{1}{2})/W$, and
$N=\lfloor{}WK/\pi\rfloor$ being the number of propagating modes in the
leads, see Eq.\ (\ref{leadspm}).
In the linear-response regime, imposed in Eq.\ (\ref{gfland}), the energy
$E$ is equivalent to the Fermi energy \cite{Naz09}; for $E>0$ the sample and
the leads show an unipolar (n-n-n) doping, for $-V_0<E<0$ we have n-p-n
structure with two interfaces separating the central region and the leads
(see Fig.\ \ref{setup:fig}).

\begin{figure}[!t]
  \includegraphics[width=\linewidth]{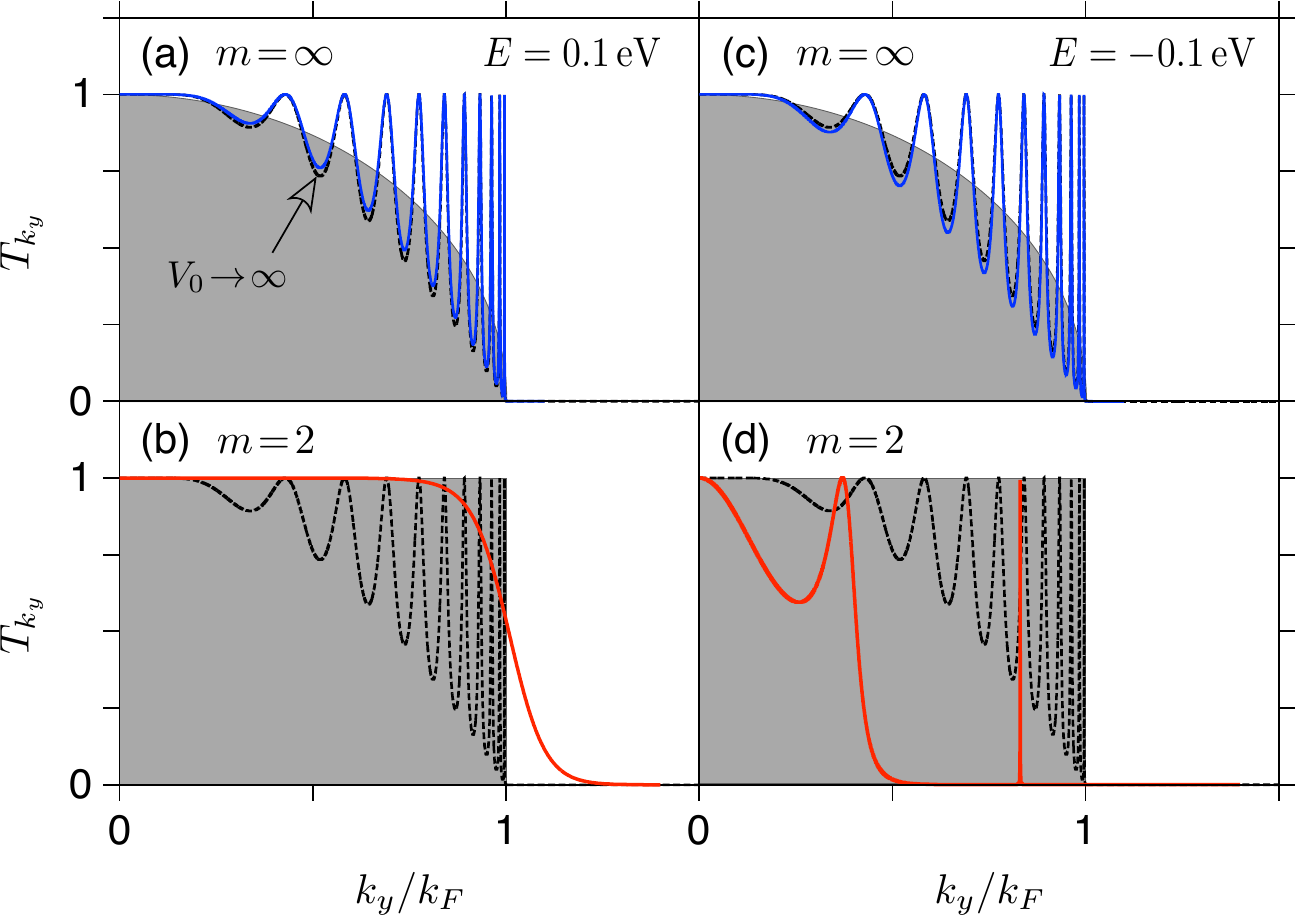}
  \caption{ \label{ttky4:fig}
    Transmission probability as a~function of the transverse momentum $k_y$
    (specified as a~fraction of $k_F=|E|/\hbar{}v_F$) at the Fermi
    energies $E=0.1\,$eV (a,b) and $E=-0.1\,$eV (c,d).
    The barrier length is $L=200\,$nm. 
    Solid lines present results obtained numerically for the potential
    given by Eq.\ (\ref{v0mpot}) with $V_0=t_0/2=1.35\,$eV and $m=\infty$
    or $m=2$ (specified at each panel).
    Dashed lines depict exact expression [see Eq.\ \ref{ttnrect}] for
    infinite rectangular barrier (a--d).
    Shaded areas mark the approximations given by Eq.\ (\ref{ttkyappx}) (a,c)
    or by the step function $T_{k_y}\approx\Theta\left(k_F-|k_y|\right)$ (b,d). 
  }
\end{figure}

\section{The rectangular barrier versus Sharvin contact}
\label{rectbar}

Before presenting the numerical results for smooth barriers, we first
recall analytic expressions for a~rectangular barrier of infinite height,
corresponding to $m\rightarrow{}\infty$, $V_0\rightarrow{}\infty$
in Eq.\ (\ref{v0mpot}).
Adapting the notation of Ref.\ \cite{Ryc09}, the transmission probability
can be written as
\begin{equation}
  \label{ttnrect}
  T_{k_y}(E) = \left[
    1+\left(\dfrac{k_y}{\varkappa}\right)^2\sin^2\left(\varkappa{}\,L\right)
  \right]^{-1}, 
\end{equation}
where
\begin{equation}
  \label{kncases}
  \varkappa = \begin{cases}
  \sqrt{k_F^2-k_y^2}, & \text{for }\  |k_y|\leqslant{}k_F, \\
  i\sqrt{k_y^2-k_F^2}, & \text{for }\  |k_y|>k_F, \\
  \end{cases}
\end{equation}
and the Fermi wavenumber $k_F=|E|/(\hbar{}v_F)$. 

In Fig.\ \ref{ttky4:fig}, we compare the results obtained from Eq.\
(\ref{ttnrect}) for the length fixed at $L=200\,$nm (black dashed lines) 
with the results of our numerical approach
[see Eqs.\ (\ref{phapri}--\ref{lsysrt})] for $V_0=t_0/2$, and $m=\infty$ or
$m=2$ (blue solid or red solid lines).
Here, continuous $k_y$ corresponds to
the $W\gg{}L$ limit. (We further limit the discussion to
$k_y\geqslant{}0$, as the mirror symmetry guarantees that $T_{k_y}=T_{-k_y}$
for any case.)
The Fermi energy is $E=0.1\,$eV 
in Figs.\ \ref{ttky4:fig}a and \ref{ttky4:fig}b, or $E=-0.1\,$eV
in Figs.\ \ref{ttky4:fig}c and \ref{ttky4:fig}d. 

In is clear from Figs.\ \ref{ttky4:fig}a and \ref{ttky4:fig}c that
the results a~finite and infinite $V_0$ are very close to each other,
as long as $|E|\ll{}V_0$. Analytic results for $V_0\rightarrow\infty$
are invariant upon the particle-hole transformation ($E\leftrightarrow{}-E$);
for a~finite $V_0$, this invariance is only approximate, since the number
of propagating modes in the leads per unit width,
$N/W \approx (E+V_0)/(\pi\hbar{}v_F)$ changes upon $E\leftrightarrow{}-E$. 
In both cases of $E>0$ and $E<0$ we observe fast, aperiodic oscillations
of $T_{k_y}$ with $k_y$ approaching $k_F$, and a~sudden decay for $k_y>k_F$, 
signalling that the role of evanescent waves is negligible
(notice that $k_F{}L\gg{}1$). 

For the case of a~finite and smooth parabolic barrier, a~striking
particle-hole asymmetry is visible, 
see Figs.\ \ref{ttky4:fig}b and \ref{ttky4:fig}d.
For $E>0$, we have a~smooth crossover from $T_{k_y}\approx{}1$ to
$T_{k_y}\approx{}0$ near $k_y=k_F$, resembling the well-known solution
for Schr\"{o}dinger electrons \cite{Kem35}.
For $E<0$, the transmission is strongly suppressed, except from the
resonances due to quasibound states \cite{Sil07}.

Let us now comment on obtaining simple, analytically tractable estimates
of the conductance and Fano factor. 

A~closer look at Eq.\ (\ref{ttnrect}) allows one to find out that,
when calculating the transport properties from Eq.\ (\ref{gfland})
for $k_F{}L\gg{}1$, summing over the modes averages out fast oscillations
originating from $\sin^2(\varkappa{}L)$, and exact transmission probability
may be approximated by 
\begin{align}
  \left(T_{k_y}\right)_{\rm approx} &=
  \frac{1}{\pi}\int_0^{\pi}\frac{d\varphi}{1+\left(k_y^2/\varkappa^2\right)
  \sin^2\varphi} \nonumber \\
  &= \sqrt{1-\left(k_y/k_F\right)^2},
  \label{ttkyappx}
\end{align}
for $|k_y|\leqslant{}k_F$; otherwise, $\left(T_{k_y}\right)_{\rm approx} =0$. 
In particular, for the conductance $G$ in the $W\gg{}L$ limit, we can put
\begin{equation}
  \label{gsubsha}
  G\approx{} \frac{g_0{}W}{\pi} 
  \int_0^{\infty}dk_y\left(T_{k_y}\right)_{\rm approx} = \frac{\pi}{4}\,G_{\rm Sharvin},
\end{equation}
with $G_{\rm Sharvin}=g_0k_F{}W/\pi$.
It is worth to notice that $G\approx{}G_{\rm Sharvin}$ corresponds to
$T_{k_y}\approx\Theta\left(k_F-|k_y|\right)$, where $\Theta(x)$ denotes
the Heaviside step function --- see shaded areas in Figs.\ \ref{ttky4:fig}(c)
and \ref{ttky4:fig}(d). 
--- representing a~reasonable approximation in the $m=2$
(parabolic barrier) and $E>0$ case, at least if one focuses on the area
under the $T_{k_y}$ plot. 
In the remaining parts of this paper, $G$ close to the approximation given
by Eq.\ (\ref{gsubsha})
is called the {\em sub-Sharvin} conductance.  

For the Fano factor, see Eq.\ (\ref{gfland}), we need to employ both
Eq.\ (\ref{ttkyappx}) and the analogous expression for $T_{k_y}^2$, namely
\begin{align}
  \left(T_{k_y}^2\right)_{\rm approx} &=
  \frac{1}{\pi}\int_0^{\pi}\frac{d\varphi}{\left[1+\left(k_y^2/\varkappa^2\right)
  \sin^2\varphi\right]^2} \nonumber \\
  &= \sqrt{1-\left(\frac{k_y}{k_F}\right)^2}
  \left[1-\frac{1}{2}\left(\frac{k_y}{k_F}\right)^2\right],   
  \label{tt2kyappx}
\end{align}
for $|k_y|\leqslant{}k_F$, or $\left(T_{k_y}^2\right)_{\rm approx} =0$
for $|k_y|>k_F$. 
In turn, we immediately obtain
\begin{equation}
  \label{ffsubsha}
  F\approx{}1-\frac{\int_0^{\infty}dk_y\left(T_{k_y}^2\right)_{\rm approx}}{
  \int_0^{\infty}dk_y\left(T_{k_y}\right)_{\rm approx}} = \frac{1}{8},
\end{equation}
constituting a~hallmark of the sub-Sharvin transport regime. 

For the sake of completeness, we also recal the results of Refs.\
\cite{Kat06,Two06} for $E=0$. 
In such a~case, Eq.\ (\ref{ttnrect}) reduces to
\begin{equation}
  T_{k_y}(0) = \frac{1}{\cosh^2\left( k_y{}L \right)}, 
\end{equation}
and integrations over $k_y$, analogous to the performed in
Eqs.\ (\ref{gsubsha}) and (\ref{ffsubsha}), leads to
\begin{equation}
  \label{gfpdiff}
  G_{\rm min} = \frac{g_0W}{\pi{}L}, \ \ \ \ \ 
  \text{and } \ \ \ \ \ 
  F_{\rm max}=\frac{1}{3}, 
\end{equation}
indicating the {\em pseudodiffusive} transport regime. (It is further
denoted in Eq.\ (\ref{gfpdiff}) that $G$ has a~minimum, whereas $F$ has
a~maximum at $E=0$.) 
The energy range $-E_{\rm diff}<E<E_{\rm diff}$, in which the pseudodiffusive
transport prevails the ballistic transport, can roughly be estimated
comparing $G_{\rm min}=G_{\rm Sharvin}(\pm{}E_{\rm diff})$, what leads to
\begin{equation}
  \label{ediff}
  E_{\rm diff} = \frac{\hbar{}v_F}{L} \approx{} 2.9\,\text{meV}\ \ \text{for }\ 
  L = 200\,\text{nm}. 
\end{equation}
The above is close to a~familiar energy uncertainty in quantum mechanics,
since the ballistic time of flight is $\Delta{}t\sim{}L/v_F$
(up to the order of magnitude).

\begin{figure}[!t]
  \includegraphics[width=\linewidth]{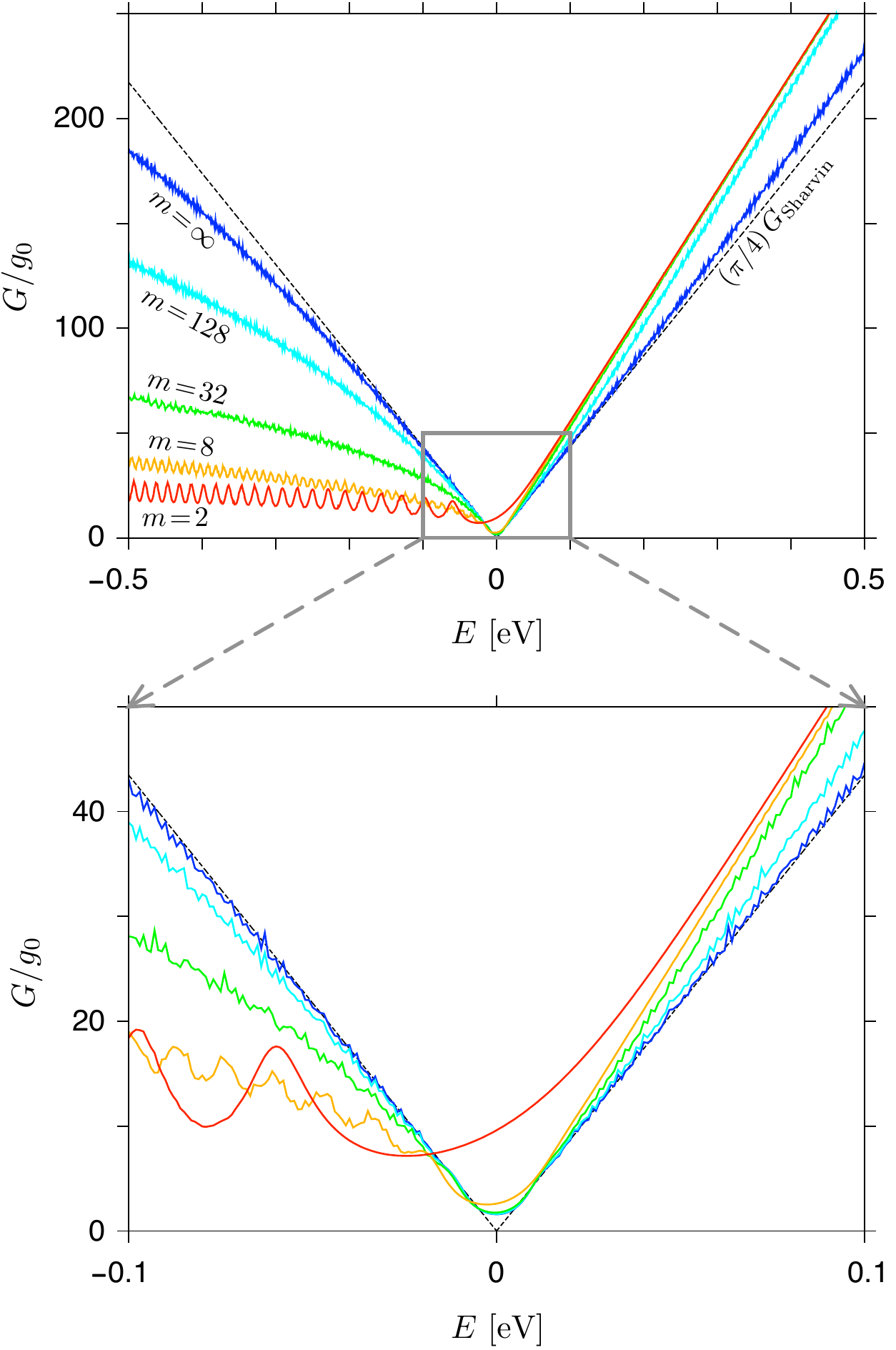}
  \caption{ \label{cond:fig}
    Conductance as a~function of the Fermi energy for the system of
    Fig.\ \ref{setup:fig}. The parameters are: $W=5\,L=1000\,$nm,
    $V_0=t_0/2=1.35\,$eV. The exponent $m$ in Eq.\ (\ref{v0mpot}) is specified
    for each dataset (solid lines). Dashed line depicts the sub-Sharvin
    conductance given by Eq.\ (\ref{gsubsha}). (The values of $G_{\rm Sharvin}$
    are not shown, as they closely follow the numerical results for $m=2$.)
    Bottom panel is zoom-in of the data presented in top panel. 
  }
\end{figure}

\begin{figure}[!t]
  \includegraphics[width=\linewidth]{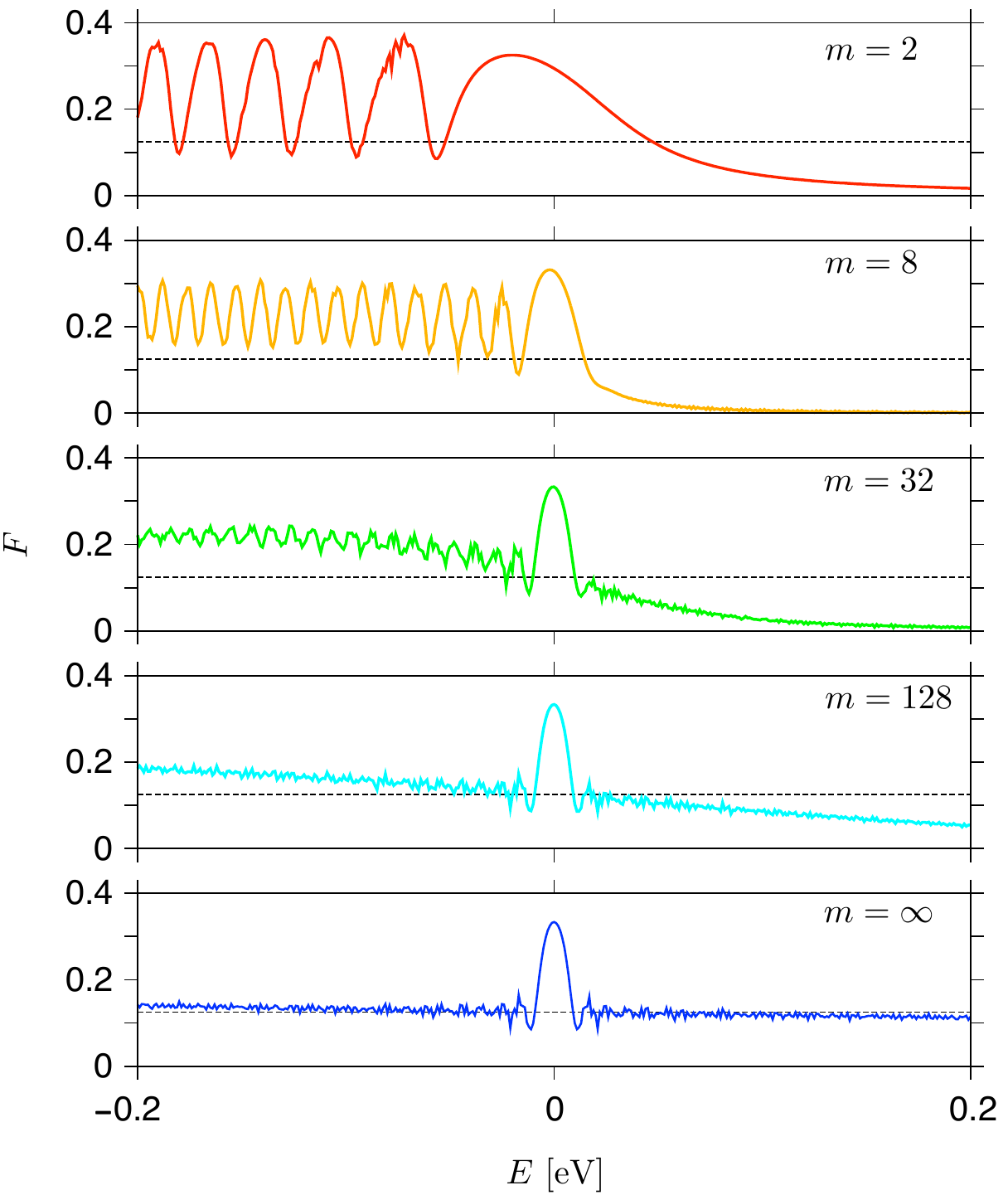}
  \caption{ \label{fano:fig}
    Fano factor as a~function of the Fermi energy for the same system
    parameters as in Fig.\ \ref{cond:fig} (solid lines).
    The exponent $m$ in Eq.\ (\ref{v0mpot}) is varied between the panels.
    Dashed line at each panel depits the sub-Sharvin value of $F=1/8$, 
    see Eq.\ (\ref{ffsubsha}). 
  }
\end{figure}

\section{Results and discussion}
\label{resudis}

In this Section, we present central results of the paper,
concerning the conductance and the Fano factor for a~graphene strip
depicted in Fig.\ \ref{setup:fig}.
The numerical calculations are carried out according to Eqs.\ 
(\ref{phapri}--\ref{gfland}), for the system with infinite-mass boundary
conditions and the width $W=5\,L=1000\,$nm. 
The step height in Eq.\ (\ref{v0mpot}) is $V_0=t_0/2$ (corresponding to 
$691\leqslant{}N\leqslant{}802$ propagating modes in the leads for
$-0.1\,\text{eV}\leqslant{}E\leqslant{}0.1\,$eV)
\cite{simfoo}.

The evolution of the conductance spectrum with the exponent $m$ in Eq.\
(\ref{v0mpot}), defining the potential profile, is visualized 
in Fig.\ \ref{cond:fig} [solid lines].
We focus now on the behavior of $G(E)$ for
$|E|\gg{}E_{\rm diff}$, see Eq.\ (\ref{ediff}), since a~close
vicinity of $E=0$ requires a~separate discussion. 

Depending on whether the system is unipolar ($E>0$) or contains
p-n junctions ($E<0$), different behaviors are observed:
For $E>0$, $G(E)$ shows a~transition, with growing $m$, from $G_{\rm Sharvin}$
to sub-Sharvin $G\approx{}(\pi/4)\,G_{\rm Sharvin}$ [dashed line].
Comparing the plots
with different energy ranges (top and bottom panel in Fig.\ \ref{cond:fig})
we immediately notice that the higher the energy, the slower convergence
with growing $m$ occurs. In fact, even the results for a~rectangular
barrier ($m=\infty$) do not match precisely Eq.\ (\ref{gsubsha}) due
to a~finite value of $V_0$. The deviations are, however, within the scale
of Fabry-Perrot oscillations, as long as $|E|\leqslant{}0.1\,$eV.
For $E<0$, the conductance is noticeably suppressed for any finite $m$,
and shows a~slow convergence (from the bottom) to sub-Sharvin values with
growing $m$. Contrary to the $E>0$ case, the values of
$G>(\pi/4)\,G_{\rm Sharvin}$ are not observed for $E<0$, except from a~close
surrounding of $E=0$. 
As a~secondary feature of the $E<0$ data, we notice relatively strong
conductance oscillations due to resonances with quasibound states. 

The above observations are further supported with the evolution of Fano
factor presented in Fig.\ \ref{fano:fig}.
Again, for $|E|\gg{}E_{\rm diff}$, the role of pseudodiffusive transport is
irrelevant and the evolution of $F$, with growing $m$, follows one of the
two distinct scenarios: For $E>0$, we have a~systematic crossover from
$F\approx{}0$ to $F\approx{}1/8$; see Eq.\ (\ref{ffsubsha}).
In contrast, for $E<0$, strong oscillation of $F$ are first suppressed
with increasing $m$, and than --- for higher $m$ --- slow convergence
of a~mean to $F\approx{}1/8$ (from the top) becomes visible. 
Similarly as for the conductance, the particle hole symmetry is only
approximate even for $m=\infty$, since the barrier height 
$V_0=t_0/2<\infty$.

\begin{figure}[!t]
  \includegraphics[width=\linewidth]{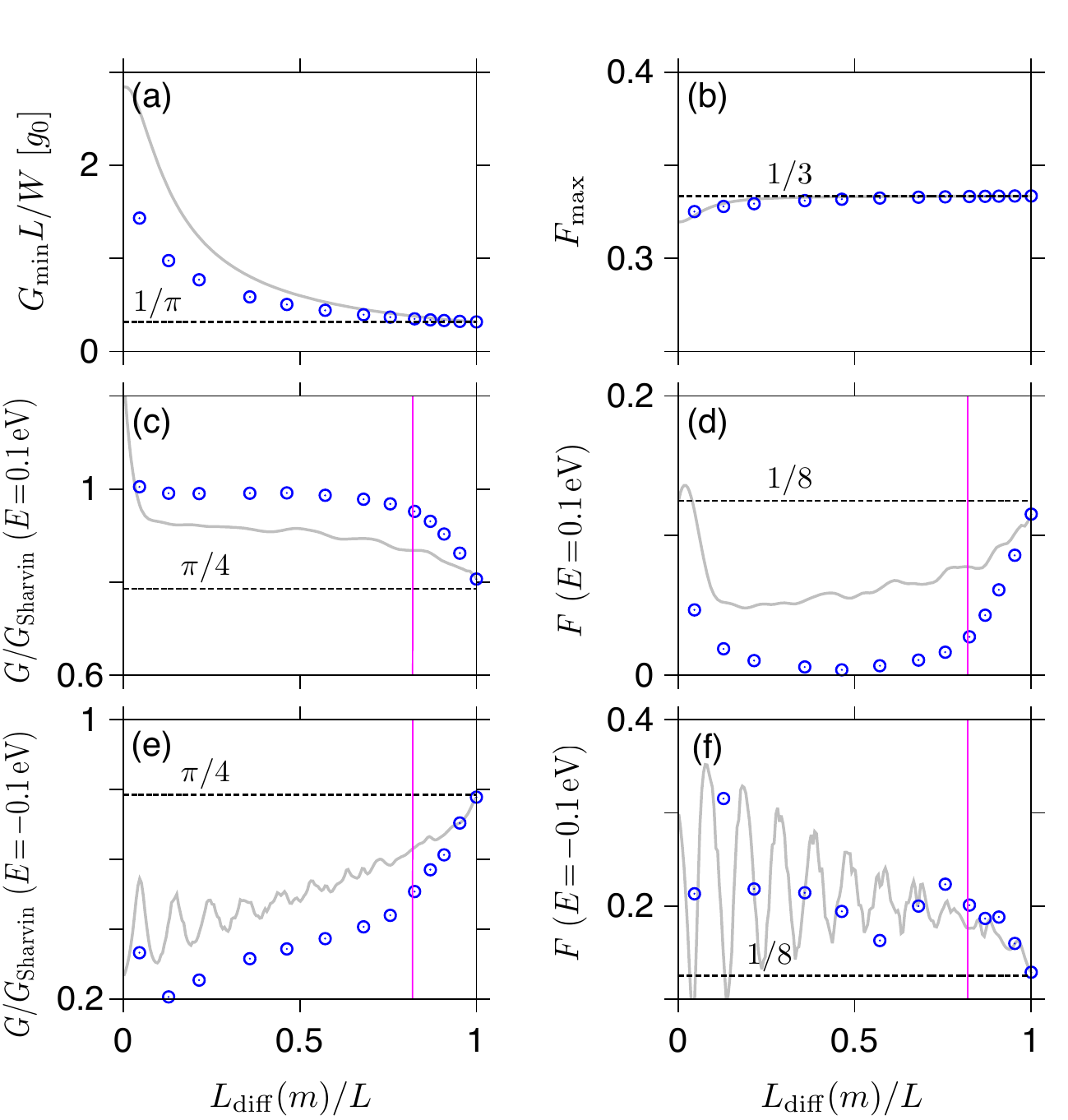}
  \caption{ \label{mima:fig}
    (a--f) Selected characteristics of the datasets presented
    in Figs.\ \ref{cond:fig} and \ref{fano:fig} displayed as functions of
    the dimensionless $L_{\rm diff}(m)/L$, see Eq.\ (\ref{ldiffm}).
    Datapoints: 
    (a) Minimal conductance. (b) Maximal Fano factor.
    (c) Conductance and (d) Fano factor at $E=0.1\,$eV.
    (e,f) Same as (c,d) but for  $E=-0.1\,$eV.
    Dashed horizontal lines depict theoretical values for
    $m\rightarrow\infty$, $V_0\rightarrow\infty$,
    see Eqs.\ (\ref{gsubsha}), (\ref{ffsubsha}), and (\ref{gfpdiff}).
    Vertical lines (c--f) mark a~bound on the right-hand side of 
    Eq.\ (\ref{ldiffbound}) for $|E|=0.1\,$eV.
    Grey solid lines (a--f) show the results for trapezoidal barrier,
    see Eqs.\ (\ref{v0trap}) and (\ref{ldiffx1}). 
  } 
\end{figure}

In order to describe the above-presented evolution of $G$ and $F$ upon
tuning the potential profile in a~quantitative manner, we display now
(in Fig.\ \ref{mima:fig}) some characteristic values extracted from
the curves in Figs.\ \ref{cond:fig} and \ref{fano:fig}.

Let us focus on the low-energy behavior of $G$ and $F$, which have not been
addressed so far in this Section. 
Using the value of $E_{\rm diff}$ given by Eq.\ (\ref{ediff}), one can define
the effective ($m$-dependent) length $L_{\rm diff}$, such that a~barrier 
can be regarded as flat for $-L_{\rm diff}/2<x<L_{\rm diff}/2$.
Requesting that $V(\pm{}L_{\rm diff}/2)=-E_{\rm diff}$, one obtains  
\begin{equation}
  \label{ldiffm}
  L_{\rm diff}(m) = L\left(\frac{\hbar{}v_F}{LV_0}\right)^{1/m}, 
\end{equation}
reducing to $L_{\rm diff}(\infty)=L$ for a~rectangular barrier.
Such a~definition allows one to present the results for any $m>0$
in a~compact range, i.e.\ $0<L_{\rm diff}(m)/L\leqslant{}1$. 

Both the minimal conductance and the maximal Fano factor, see Figs.\
\ref{mima:fig}(a) and \ref{mima:fig}(b), show rapid convergence (with
$L_{\rm diff}/L\rightarrow{}1$) to the values predicted for infinite rectangular
barrier, see Eq.\ (\ref{gfpdiff}). In particular, $F_{\rm max}\approx{}0.325$
for the lowest considered $m=2$ (corresponding to $L_{\rm diff}/L\approx{}0.046$).
This finding illustrates how the precise value of $F_{\rm max}$ is insensitive
to the details of electrostatic potential profile, helping to understand,
why experimental values of $F$ are sometimes surprisingly close to $1/3$ 
\cite{Dan08,Dic08,Kum15}. 

Away from the charge-neutrality point, the system characteristics presented
in Figs.\ \ref{mima:fig}(c)--\ref{mima:fig}(f) show a~different behavior. 
Namely, they generally take values rather distant from predictions given
(respectively) in Eqs.\ (\ref{gsubsha}) and (\ref{ffsubsha}), except from
a~relatively narrow interval near $L_{\rm diff}/L=1$, in which systematic
convergence occurs. 

A~brief explanation is provided below. 

For $|E|\gg{}E_{\rm diff}$, the evolution of $G$ and $F$ depends on a~mutual
relation between the Fermi wavelength $\lambda_F(E)=2\pi/k_F=h{}v_F/|E|$,
and the characteristic lenghtscale of a~potential jump, which can be
defined as $\Delta{}x = (L-L_{\rm diff})/2$, with $L_{\rm diff}$
given by Eq.\ (\ref{ldiffm}). If
\begin{equation}
  \label{lamfdx}
  \lambda_F/2\lesssim\Delta{}x,
\end{equation}
the barrier
cannot longer be regarded as (even approximately) rectangular.
Since $\Delta{}x$ is related to $m$ via $L_{\rm diff}$, Eq.\ (\ref{lamfdx}) 
can be rewritten as
\begin{equation}
  \label{ldiffbound}
  \frac{L_{\rm diff}}{L}\lesssim{} 1-\frac{2\pi\hbar{}v_F}{L|E|},  
\end{equation}
giving $L_{\rm diff}/L\lesssim{}0.82$ for $|E|=0.1\,$eV. 
The structure of Eq.\ (\ref{ldiffbound}) guarantees that the inequality is
always satisfied for $L_{\rm diff}<L$ (i.e., the smooth potential, $m<+\infty$)
and sufficiently high $|E|$. In such a~case, the measurable quantities
deviate from the predictions for a~rectangular barrier. 

Once the upper bound in Eq.\ (\ref{ldiffbound}) marked with vertical line
in Figs.\ \ref{mima:fig}(c)--\ref{mima:fig}(f) is exceeded (i.e.,
$\lambda_F>2\Delta{}x$), a~systematic convergence of all the considered
quantities, with $L_{\rm diff}/L\rightarrow{}1$, to the predictions
for a~rectangular barrier becomes visible. 

Finally, it is worth to compare our results with those corresponding for
trapezoidal barrier, discussed in Ref.\ \cite{Par21}. The electrostatic
potential energy can be written as 
\begin{equation}
  \label{v0trap}
  V(x) = -V_0\times
  \begin{cases}
    \,0  &  \text{if }\ |x| \leqslant x_1, \\
    \,{\displaystyle\frac{|x|-x_1}{x_0-x_1}}
    &  \text{if }\ x_1<|x|\leqslant x_0, \\
    \,1  &  \text{if }\ |x| > x_0, 
  \end{cases}
\end{equation}
with $0 \leqslant x_1  \leqslant x_0$, parametrizing the barrier evolution
between the limiting cases of triangular ($x_1=0$) and rectangular
shape ($x_1=x_0$). In analogy to Eq.\ (\ref{ldiffm}), we have
\begin{equation}
  \label{ldiffx1}
  L_{\rm diff}(x_1) = 2x_1 + \frac{\hbar{}v_F}{LV_0}\left(L\!-\!2x_1\right)
  \approx{}2x_1. 
\end{equation}

Taking the same values of $W$, $L$, and $V_0$ as before, and varying $x_1$,
one can easily find that the conductance and Fano factor spectrum evolves
in a~qualitatively similar manner as the spectra depicted in
Figs.\ \ref{cond:fig} and \ref{fano:fig}.
Several quantitative differences can be identified, however, referring
to the numerical characteristics presented in Fig.\ \ref{mima:fig}
[grey solid lines].
First, for $L_{\rm diff}\ll{}L$ and $E>0$, the conductance significantly
exceeds the value of $G_{\rm Sharvin}$, and the Fano factor is also enhanced
compared to the smooth potentials;  
see Figs.\ \ref{mima:fig}(c) and \ref{mima:fig}(d).
Most remarkably, the values of $G\approx{}G_{\rm Sharvin}$ and $F\approx{}0$
are never approached for trapezoidal potentials, showing that standard
ballistic transport may be restored in bulk graphene ($W\gg{}L$) only
for smooth barriers. 
For $E<0$, the behavior of $G$ and $F$ is similar for smooth and trapezoidal
potentials, see Figs.\ \ref{mima:fig}(e) and \ref{mima:fig}(f); some
enhancement of $G$ (and slightly faster convergence to sub-Sharvin value
with $L_{\rm diff}/L\rightarrow{}1$) can be noticed for trapezoidal barriers.

\section{Conclusions}
\label{conclus}

We have identified sub-Sharvin transport regime in ballistic graphene,
which manifests itself via the suppressed conductivity, $G\approx
(\pi/4)\,G_{\rm Sharvin}$ (with $G_{\rm Sharvin}=g_0k_F{}W/\pi{}$, $g_0=4e^2/h$ the
conductance quantum, $k_F$ the Fermi wavenumber, and $W$ the sample width),
and the enhanced shot noise, $F\approx{}1/8$, comparing to standard 
quantum point contacts. Solving the scattering problem numerically, for
different electrostatic potential profiles, we find that such a~regime appears
generically for rectangular and smooth potential barriers, provided that
the following conditions are satisfied:
(i) the sample width $W\gg{}L$, with $L$ the sample length,
(ii) the Fermi wavenumber $k_F{}\gg{}L^{-1}$, and (iii) the Fermi wavelength
$\lambda_F=2\pi/k_F\gg{}\Delta{}x$, with $\Delta{}x$ being the linear size
of an interface between weakly- and heavily-doped graphene areas (i.e., the
sample and the leads).

Taking into account that highest accessible Fermi
energies in electrostically-doped graphene devices are 
$E=\pm{}\hbar{}v_Fk_F\approx\pm{}0.1\,$eV, condition (iii) is equivalent to
$\Delta{}x\ll{}36\,$nm, showing that atomistic precision in tailoring
the spatial potential profile is not necessary to detect the effects we
describe. Moreover, for $\lambda_F\gtrsim{}2\Delta{}x$ (being equivalent to
$k_F\lesssim{}\pi/\Delta{}x$), we predict
a~monotonous convergence, with increasing $\lambda_F$ (or shrinking
$\Delta{}x$), of the transport characteristics to the values expected
for the sub-Sharvin regime. 

Our results thus complement previous studies (see Refs.\
\cite{Par21,Sil07,Che06})
in which the range of $W\gg{}L$ and $L^{-1}\ll{}k_F\lesssim{}\pi/\Delta{}x$
have not been elaborated.
In such a~range, a~family of smooth barriers considered here 
leads to clear crossover (for electronic dopings) from Sharvin to
sub-Sharvin transport regime upon tuning the barrier shape, with
Sharvin characteristics occurring in a~considerable range of steering
parameters. 
This feature is absent for trapezoidal barriers proposed in 
Ref.\ \cite{Par21}. 
Since the carrier diffusion in real device must lead to the effective
potential varying smoothly in an interface between areas of different dopings,
we believe the above mentioned crossover should be observable.

Although the present work focusses on graphene, we expect the main effects
to reappear in other two-dimensional systems such as silicene
\cite{Eza15}, since the sub-Sharvin transport is link to conical dispersion
relation rather then to the transmission via evanescent waves (responsible for
graphene-specific phenomena occuring at the charge-neutrality point). 

\section*{Acknowledgments}
The work was supported by the National Science Centre of Poland (NCN)
via Grant No.\ 2014/14/E/ST3/00256.



\begin{thebibliography}{}
\bibitem{leadfoo}
Comparing the results with these for a~rectangular barrier and heavily-doped
leads ($m\rightarrow\infty$, $V_0\rightarrow\infty$) we replace Eq.\
(\ref{leadspm}) with $\phi_\infty^{(\pm)}=(1,\pm{}1)^T{}e^{\pm{}iK_x{}x}$.
\bibitem{simfoo}
Numerical integration of Eqs.\ (\ref{phapri},\ref{phbpri})
were performed utilizing a~standard forth-order Runge-Kutta
algorithm. A~spacial step of $\Delta{}x=5\,$pm was sufficient to keep
the unitarity error, $\mbox{max}\left(|\epsilon_n|\right)<10^{-8}$,
with $\epsilon_n=|r_n|^2+|t_n|^2-1$. 
Summation over the modes in Eq.\ (\ref{gfland}) was terminated
if $T_n<10^{-6}$.
\end{thebibliography}
\end{document}